\documentstyle[12pt,aps]{revtex}
\begin{document}

\title{
\hfill\parbox[t]{2in}{\rm\small\baselineskip 14pt 
JLAB-TH-96-16 }
\vskip 1.3cm 
Exclusive versus Inclusive Semileptonic $\bar B$ Decays\\ in the Quark Model\\~
}

\vspace {0.5 cm}

\author{Nathan Isgur}
\address{Jefferson Lab, 12000 Jefferson Avenue, Newport News, Virginia  23606\\~}

\vspace {0.5 cm}

\maketitle

%%%%%%%%%%%%%%%%%%%%%%%%%%%%%%%%%%%%%%%%%%%%%%%%%%%%%%%%%%%%%%%%%%%%
%\begin{center}  {\bf Abstract}  \end{center}
%%%%%%%%%%%%%%%%%%%%%%%%%%%%%%%%%%%%%%%%%%%%%%%%%%%%%%%%%%%%%%%%%%%%

\begin{abstract}

Some emerging difficulties in the theoretical description of exclusive 
semileptonic $\bar B$ decays are discussed in the context of the quark model. 
While there are no unambiguous problems at this time, I discuss physics 
beyond the valence quark model which should eventually be 
probed by precision measurements of $\bar B$ semileptonic decays.
\vspace {0.5 cm}

\end{abstract}

   Wolfenstein [1] has commented on an emerging discrepancy between the measured rate
of inclusive semileptonic $\bar B$ decay and the sum of the rates to the exclusive channels
considered in the Isgur-Scora-Grinstein-Wise (ISGW) quark model [2]. 
While calling attention to this issue is very valuable, 
I disagree with Wolfenstein's interpretation of its implications. 
In particular,
I will argue that if there is rate missing from the sum over
exclusive channels, then the most likely origins are
nonresonant decays and highly excited resonances that lie outside the scope of the ISGW
model, and not in a problem with the model itself.

   We should begin these considerations by recognizing
that the ISGW quark model should not in general be expected to be able to make predictions with
better than typical quark model accuracy since, among other things,
it is grounded in the $1/N_c$ expansion, so it assumes valence quark dominance, and
while it respects relativistic kinematics, 
it calculates the form factors for semileptonic 
decays using nonrelativistic 
valence quark wavefunctions.
At the same time, we note that in its updated version as ISGW2 [2], 
this model respects the constraints
of Heavy Quark Symmetry [3] and so in some cases its model-dependence appears
only in $1/m_Q$ terms.

   Let me next address the issue of the theoretical
consistency between the ISGW2 model and
QCD-corrected inclusive 
$b \rightarrow c \ell \bar \nu_{\ell}$ calculations. The latter calculations give 
$\Gamma_{sl}=(4.6 \pm 0.3) \vert V_{cb} \vert ^2 \times 10^{13}$; the  theoretical
error I have assigned to this result 
will be discussed below.  
ISGW2 gives 
$\Gamma(\bar B \rightarrow D \ell \bar \nu_{\ell})=1.2 ~\vert V_{cb} \vert ^2 \times 10^{13}$,
$\Gamma(\bar B \rightarrow D^* \ell \bar \nu_{\ell})=2.5~ \vert V_{cb} \vert ^2 \times 10^{13}$,
and a rate to the three lowest-lying excited heavy quark spin multiplets 
with $s_{\ell}^{\pi_{\ell}}= {1\over 2}^-$, ${3\over 2}^-$, and ${1\over 2}^+$ 
of $0.4~ \vert V_{cb} \vert ^2 \times 10^{13}$. These exclusive modes correspond
to $26 \pm 2\%$, $54\pm 4\%$, and $8 \pm 1\%$ of $\Gamma_{sl}$ leaving 
$12 \pm 6\%$ of the rate unaccounted for {\it theoretically}.

   Note that the $1/N_c$ valence approximation is irrelevant to the 
issue of the consistency between ISGW and inclusive
calculations since within that approximation a
complete exclusive calculation and the inclusive calculation should agree.  
So where is the missing rate?  It can be in three places:

\medskip

1.  Without explicitly calculated matrix elements to yet more highly excited states,
ISGW is unable to quantitatively address the completeness of their truncated 
sum over exclusive channels for 
$b \rightarrow c \ell \bar \nu_{\ell}$ transitions. However, from the convergence 
they see with excitation energy in $\bar B$ decays 
and the increasing shortfall with respect to the inclusive rate they see 
in $\bar B_s$ and $\bar B_c$, it would not be surprising if the $\bar B$ decay
rate to all yet higher 
spin multiplets were equal to that to the three excited spin multiplets
they explicitly compute, namely about another 
$8\%$.  If so, 
the exclusive-inclusive discrepancy would be an insignificant $4 \pm 6\%$. Note 
that the rate of convergence of the sum over exclusive channels is controlled by
how close $b \rightarrow c \ell \bar \nu_{\ell}$ decays are to the Shifman-Voloshin
limit [4].

2.  The inclusive rates have explicit QCD radiation in them.  Such radiation 
is consistent with the $1/N_c$ valence approximation, but corresponds to 
the excitation of hybrid mesons which are ignored in ISGW.  From the contribution 
of radiative corrections to the recoil dependence
of the $D$ and $D^*$ rates, one can estimate using  Bjorken's sum 
rule [5,6] about a $4\%$ contribution of such states.  The exclusive-inclusive 
discrepancy would now be $0 \pm 6\%$.

3.  The reliability of the inclusive rate calculation is still unclear.  
The theoretical error we have assigned was intended to be 
adequate to cover the uncertainty in QCD radiative
corrections, but the total error could be considerably larger given how incompletely 
$1/m_Q$ effects (associated with both mass shifts $\bar m_B=m_b+\bar \Lambda$ and 
the accuracy of quark-hadron duality) are understood [7].

\medskip

\noindent In summary, there is no clear indication that the ISGW model 
is theoretically inconsistent as gauged by its correspondence to inclusive calculations.

   Let us now turn to the experimental situation. We first note that 
experiment [8] gives $D$ and $D^*$ semileptonic rates of $19\pm 5\%$ and $45\pm 3\%$,
each somewhat smaller than the ISGW2 predictions. Wolfenstein focuses 
on the fact that these measurements imply that
$36\pm 6\%$ of the rate goes to other states, 
versus the $8\pm 1\%$ explicitly taken into account by ISGW2. Based on the preceeding discussion, 
one could instead take the point of view that ISGW2 expected $20\pm 6\%$ of the decays to be to
excited states (a $2 \sigma$ discrepancy), and that it
explicitly calculated the rate to about half of these excited state decays.

   Recent experimental findings lend support to this view.
Wolfenstein's Comment depends to some extent on the 1995 publication
by the OPAL collaboration [9] 
reporting very large branching ratios to the $D_1(2420)$ and $D^*_2(2460)$ states 
of the $s_{\ell}^{\pi_{\ell}}= {3\over 2}^-$ multiplet.  
These reports, if confirmed, would 
have neatly accounted for the ``missing" $36 \pm 6\%$ of the semileptonic rate.  However, such 
a large strength to those states seemed to be in conflict with the observed [10]
slope $\rho^2=0.84 \pm 0.14$ of the Isgur-Wise function, which strongly suggests
{\it via} Bjorken's sum rule 
a much smaller $s_{\ell}^{\pi_{\ell}}= {3\over 2}^-$ strength closer to that 
of ISGW2 (where $\rho^2=0.74$).  Recent measurements have indeed changed matters 
substantially:  ALEPH [11] reports $7 \pm 2\%$ of the semileptonic rate to the $D_1(2420)$ 
and CLEO [12] reports $< 9\%$ at the 90\% confidence limit, to be compared 
to OPAL's $20\pm 6\%$. Moreover, measurements [8] of the decay
$\bar B \rightarrow D_1(2420) \pi$, coupled with the apparent
validity of factorization for such decays, would imply
a semileptonic $D_1(2420)$ fraction of $5\pm 2\%$. Thus the ISGW2 prediction
that this fraction is $4\%$ does not seem to be far off the mark.
For the $D_2^*(2460)$,
ALEPH reports $< 4\%$ at the 90\% 
confidence limit to be compared to OPAL's $22\pm 9\%$. ISGW2 predicts this rate 
to be 2\%.  At the same time, ALEPH reports that the final states
$D \pi \ell \bar \nu_{\ell}$ and
$D^* \pi \ell \bar \nu_{\ell}$ account 
for $21\pm 5\%$ of the $36\pm 6\%$ of the $\bar B$ semileptonic rate that was 
not $D$ or $D^*$.  Recall that ISGW2 has $20\pm 6\%$ non-$D+D^*$ decays, of 
which $8\pm 1\%$ is in explicitly summed channels.
The ALEPH observations are thus consistent with ISGW2 if it is indeed the case
that $12\pm 6\%$ of the 
semileptonic decays go into highly excited $D$ mesons (both quarkonia and hybrids).
I would conclude that it is premature to declare that there is a serious discrepancy
between ISGW2 {\it per se} and experiment.

   I would nevertheless like to agree with Wolfenstein
that there {\it are} probably more than just the ISGW2 processes contributing
to the inclusive rate. We have indeed already seen that
theoretical consistency requires $12\pm 6\%$ more rate, and have
identified  highly excited $D$ mesons not in ISGW2 as certain
sources of uncalculated rate. However, there are both theoretical and
experimental indications that nonresonant processes, which are
outside of ISGW2 since they correspond to $N_c^{-1}$ effects,
may be at least as important as these uncalculated parts
of  processes that are of leading order in $N_c$.

  As a prelude to discussing nonresonant processes, we note 
that there are, in addition to direct measurements [10],
many indirect indications that
the  prediction of ISGW2 for $\rho^2$ is too small:
the predicted $\bar B \rightarrow D  \ell \bar \nu_{\ell}$ and
$\bar B \rightarrow D^* \ell \bar \nu_{\ell}$ rates are somewhat too high,
the predicted production of all excited states is somewhat too low, and
ISGW2 predicts all of the measured analogs to $\rho^2$, namely the 
form factor slopes for $\pi \rightarrow \pi$, $K \rightarrow \pi$ and 
$D \rightarrow K$ transitions, to be too small by about 30\% [2].
These experimental problems are all consistent with 
an acknowledged [2] theoretical defect of
ISGW: its neglect of nonvalence effects. This defect can be addressed by 
``unquenching the quark model" [13], {\it i.e.} by turning on the effects of 
$q \bar q$ pairs (or equivalently of a complete set of
meson loop graphs).  When the $b$ quark decays 
from a $b \bar q q \bar q$ configuration inside the $\bar B$, it 
simply makes a corresponding configuration of the $D$ or $D^*$ at $w=1$ (in
the Heavy Quark Limit), 
but as $w-1$ is increased such configurations make increasingly small contributions 
to ``elastic" scattering relative to the $b \bar q$ configuration.  
{\it I.e.}, they will make a net positive contribution to $\rho^2$ after 
renormalization.   By  Bjorken's sum rule, this contribution will be 
dual not to the production of the $c \bar q$ resonances, but rather to a 
$c\bar q+q\bar q$ continuum.  In such an ``unquenched" version of
ISGW one would in fact naturally expect an additional contribution of order   
$10\%$ to the semileptonic rate from nonresonant states corresponding
to a conjectured $30\%$ increase in  $\rho^2$.
With additional $c \bar q$ excited states and hybrids as well as 
such nonresonant decays, the total rate to exclusive excited states could easily be of order
$30\%$.

In summary, we believe the foregoing suggests that careful study of $\bar B$ semileptonic 
decays could answer some old and very important physics
questions concerning quark-hadron duality.  To extract this 
physics, it will be important to have more accurate measurements 
of the ``elastic" $D$ and $D^*$ fractions, but especially to delineate the 
strength and nature of the non $D+D^*$ contributions.  We anticipate 
not only somewhat more resonant strength, but also a substantial 
nonresonant continuum.  Theoretically, these latter decays appear to 
provide a clear testing ground for the accuracy of the valence 
approximation.  In particular, the large energy release in a $b \rightarrow c$ 
transition will allow a probe of the non-valence components of 
the ``brown muck" out to high relative momentum.

\bigskip\bigskip\bigskip

\noindent References
\medskip

\noindent[1] L. Wolfenstein, Phys. Rev. D, to appear as a Comment.

\noindent[2] D.~Scora
and N.~ Isgur,  Phys. Rev. D{\bf 52}, 
2783 (1995) present an updated version of the ISGW model called
ISGW2. For the original paper, which includes a more complete discussion of
its foundations in the $1/N_c$ expansion and in particular 
of the possible role of nonresonant states, see
N.~Isgur, D.~Scora, B.~Grinstein,
and M.~B. Wise,  Phys. Rev. D{\bf 39}, 799 (1989); B.~Grinstein, M.~B. Wise, and N.~Isgur,
Phys. Rev. Lett. {\bf 56}, 298 (1986).

\noindent[3] N. Isgur and M.B. Wise, Phys.
Lett. {\bf B232} (1989) 113; Phys. Lett.
{\bf B237} (1990) 527. For an overview
of Heavy Quark Symmetry see N. Isgur
and M.B. Wise, ``Heavy Quark Symmetry"
in {\it B Decays}, ed. S. Stone (World
Scientific, Singapore, 1991), p. 158,
and in {\it ``Heavy Flavors"}, ed. A.J.
Buras and M. Lindner (World Scientific,
Singapore, 1992), p. 234.

\noindent[4] M.~B.~Voloshin and M.~A.~Shifman, Yad. Fiz. {\bf 47}, 801 (1988); Sov. J. Nucl.
Phys. {\bf 47}, 511 (1988); M.~A.~Shifman in {\em Proceedings of the 1987
International Symposium on Lepton and Photon Interactions at High Energies},
Hamburg, West Germany, 1987, edited by W.~Bartel and R.~R{\"u}ckl, Nucl.
Phys. B (Proc. Suppl.) {\bf 3}, 289 (1988); N. Isgur, Phys. Rev. D{\bf 40}, 109 (1989).

\noindent[5]  J.D. Bjorken, in Proceedings of
the $4^{th}$  Rencontre de
Physique de la Vallee d'Aoste, La
Thuile, Italy, 1990, ed. M. Greco
(Editions Frontieres, Gif-sur-Yvette,
France, 1990);
J.~D. Bjorken, {\em Recent Developments in Heavy Flavor Theory},
in Proceedings of the XXVth
International Conference on High Energy
Physics, Singapore, (World Scientific, Singapore, 1992).

\noindent[6]  N.~Isgur and M.~B. Wise,
Phys. Rev. D {\bf 43}, 819 (1991).

\noindent[7] See, e.g., A.F. Falk, M. Luke, and M.J. Savage, Phys. Rev. D{\bf 53},
6316 (1996) and references therein.

\noindent[8] Semileptonic B decays are  
summarized 
by the Particle Data Group,
Phys. Rev. D{\bf 54}, 1 (1996). We also note that 
I. Dunietz has suggested that the exclusive branching ratios may need revision since
$D \rightarrow K \pi$, which normalizes 
these ratios, may be about 15\% smaller than currently thought (FERMILAB-PUB-96/104-T).

\noindent[9] R. Akers {\it et al.} (The OPAL Collaboration), Z. Phys. {\bf C67}, 57 (1995).

\noindent[10] B. Barish {\it et al.}. Phys. Rev. D{\bf 51}, 1014 (1995).

\noindent[11] S. Armstrong for the ALEPH Collaboration, as reported 
at XIV International Conference on Particles and Nuclei (PANIC '96) in Williamsburg,
Virginia, May, 1996.

\noindent[12] J.P. Alexander {\it et al.} (The CLEO Collaboration), CLEO CONF 95-30 (1995).

\noindent[13] P. Geiger and N. Isgur, Phys. Rev. D {\bf41}, 1595 (1990);  
P. Geiger and N. Isgur, Phys. Rev. D {\bf44}, 799 (1991); 
Phys. Rev. Lett. {\bf67}, 1066 (1991); 
Phys. Rev. D {\bf47}, 5050 (1993);
P. Geiger, {\it ibid.} {\bf49}, 6003 (1993); P. Geiger and N. Isgur, CEBAF-TH-96-08.

\end{document}